\documentclass[twocolumn,prb,amsmath,amssymb]{revtex4}

\usepackage{graphicx}

\usepackage{dcolumn}

\usepackage{bm}

\begin{document}

%\preprint{APS/123-QED}

\title{Room temperature spin coherence in ZnO}

\author{S. Ghosh}
\author{V. Sih}
\author{W. H. Lau}
\author{D. D. Awschalom}
\email{awsch@physics.ucsb.edu}

\affiliation{Center for Spintronics and Quantum Computation,
University of California, Santa Barbara, CA 93106}

\author{S.-Y. Bae}
\author{S. Wang}
\affiliation{Department of Materials Science and Engineering,
Stanford University, Stanford, CA 94305}
\author{S. Vaidya}
\author{G. Chapline}
\affiliation{Lawrence Livermore National Laboratory, Livermore, CA
94550}

\date{\today}

\begin{abstract}

Time-resolved optical techniques are used to explore electron spin
dynamics in bulk and epilayer samples of $\it{n}$-type ZnO as a
function of temperature and magnetic field. The bulk sample yields
a spin coherence time ${\it T}_{2}^*$ of 20 ns at $\it{T}$ = 30 K.
Epilayer samples, grown by pulsed laser deposition, show a maximum
${\it T}_{2}^*$ of 2 ns at $\it{T}$ = 10 K, with spin precession
persisting up to $\it{T}$ = 280 K.
\end{abstract}

\maketitle

A lot of attention has been focused on zinc oxide (ZnO) because of
material properties that make it well-suited for applications in
ultra-violet light emitters, transparent high-power electronics
and piezoelectric transducers. In addition, the theoretical work
of Dietl $\it{et}$ $\it{al}$.,\cite{deitl} predicting room
temperature ferromagnetism for Mn-doped $\it{p}$-type ZnO, has
revealed the possibility that ZnO may be an appropriate candidate
for spintronics.\cite{wolf} The magnetic properties of thin films
of ZnO with transition ion doping,\cite{pearton,jung,fukumura} are
being widely investigated, but practical spintronics applications
would also require long spin coherence time and spin coherence
length.\par

In this letter, we investigate the electron spin dynamics of
non-magnetically doped $\it{n}$-type ZnO. The growth of
$\it{p}$-type films is an experimental challenge, possibly due to
self-compensation from oxygen vacancies\cite{look} or the
incorporation of hydrogen as an unintentional donor.\cite{chris}
Consequently, recent developments reporting magnetic properties in
ZnO have been regarding $\it{n}$-type samples,\cite{ueda} although
the preparation of $\it{p}$-type films has been
reported.\cite{minegishi,vaithianathan} Our measurements on
$\it{n}$-type ZnO establish a spin coherence time of $\sim$190 ps
at room temperature, longer than the spin coherence time reported
in GaN,\cite{beschoten} another wide band-gap semiconductor. ZnO
also has the added advantage that high quality single crystals are
commercially available.\par

We concentrate on four samples - three 100 nm epitaxial thin films
(designated samples A - C) and a bulk sample D. The thin films are
fabricated by a pulsed laser deposition system using a ceramic ZnO
pellet (Praxair Targets, Inc.) as the target and $\it{c}$-cut
sapphire as the substrate, which is heated to 800$^{o}$C during
deposition.  The substrate $\it{ab}$ plane is rotated by 30
degrees with respect to that of the epilayers to reduce the
lattice mismatch from 16 $\%$ to 3.85 $\%$. Four-circle X-ray
diffractometry confirms the single phase growth and the wurtzite
structure of ZnO. The oxygen pressure used during the growth
process for the epilayers is 10$^{-5}$ torr (sample A), 10$^{-4}$
torr (B) and 10$^{-3}$ torr (C), which tunes the carrier
concentration. Sample D (commercial single crystal from
Tokyo-Denpa Company Ltd. grown by hydrothermal
method\cite{sakagami}) is mounted on fused silica and polished
down to a few microns for transmission measurements. \par

Transport measurements performed at room temperature show
$\it{n}$-type conductivity in all four samples. The carrier
concentrations (mobilities) for A - D are 1.92 x 10$^{19}$
cm$^{-3}$ (19 cm$^{2}$/Vs), 1.01 x 10$^{19}$ cm$^{-3}$ (27
cm$^{2}$/Vs), 2.64 x 10$^{18}$ cm$^{-3}$ (37 cm$^{2}$/Vs), and
1.26 x 10$^{15}$ cm$^{-3}$ (240 cm$^{2}$/Vs), respectively. We
note that sample D has a carrier concentration several orders of
magnitude smaller than the epilayers and a higher mobility.\par

Time-resolved Faraday rotation (TRFR), an optical pump-probe
spectroscopic technique,\cite{crooker1,crooker2} is used to probe
the electron spin dynamics. A circularly-polarized pump pulse,
incident normal to the sample surface, injects spin-polarized
electrons, and the Faraday rotation angle of a linearly polarized
probe pulse, applied after a time delay $\it{\Delta t}$, measures
the projection of the electron spin magnetization as it precesses
in a plane perpendicular to the applied transverse magnetic field
(Voigt geometry). The pump and probe are typically tuned to 369 nm
to address the band-gap of ZnO, using the frequency-doubled output
from a mode-locked Ti:Sapphire laser with pulse duration $\sim$
150 fs and repetition rate of 76 MHz ($\it{t_{rep}}$ = 13 ns). The
laser beams are focused to a spot size of $\sim$ 50 $\mu$m, and
typical pump and probe powers are 1.9 mW and 200 $\mu$W,
respectively. The circular polarization of the pump beam is
modulated with a photoelastic modulator at 50 kHz for lock-in
detection. \par

\begin{figure}
\includegraphics[width=0.44\textwidth]{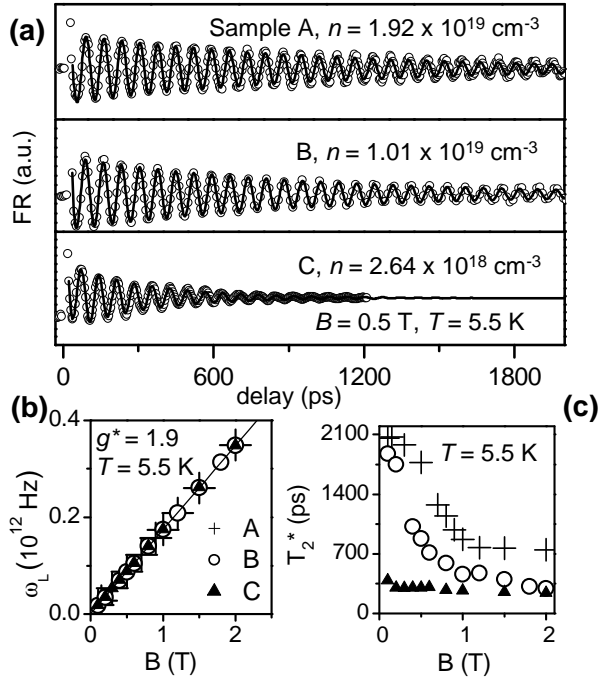}
\caption{\label{fig:epsart}(a) Time-resolved Faraday rotation as a
function of pump-probe delay for samples A - C at $\it{T}$ = 5.5 K
and $\it{B}$ = 0.5 T. The carrier densities are indicated in the
figure. Lines are fits using eq. (1). (b) Larmor frequency
$\omega_{L}$ as a function of $\it{B}$. Line is a linear fit. (c)
Spin coherence time $\it{T}_{2}^{*}$ as a function of $\it{B}$.
The symbols in (c) correspond to those in (b).}
\end{figure}

TRFR results for samples A - C at T = 5.5 K and B = 0.5 T are
shown in Fig. 1. The oscillatory component of the Faraday rotation
$\theta_{F}$ as a function of time delay is:
\begin{equation}
\theta_{F}(\Delta t)  = Aexp(-\Delta
t/T_{2}^{*})cos(\omega_{L}\Delta t)
\end{equation}
where $\it{A}$ is the amplitude of the electron spin polarization
injected perpendicular to $\it{B}$, ${\it T}_{2}^{*}$ the
transverse coherence time and $\omega_{L}$ the Larmor frequency
which is related to the electron g-factor $\it{g^{*}}$ by
$\hbar\omega_{L}$ = $\it{g^{*}}\mu_{B}\it{B}$ (where $\mu_{B}$ is
the Bohr magneton). Fits to the data in Fig. 1(a) indicate that
${\it T}_{2}^{*}$ increases with increasing carrier concentration
in the epilayers, from 0.5 ns (sample C) to 2.1 ns (sample A).\par

Figures 1(b) and (c) show the $\it{B}$ - dependence of
$\omega_{L}$ and $\it{T_{2}^{*}}$, respectively. In Fig. 1(b),
$\omega_{L}$ increases linearly with $\it{B}$ in all the samples
(including sample D, not shown here) and we extract a
field-independent value of $\it{g^{*}}$ = 1.9 for the three
epilayers, consistent with earlier electron spin resonance (ESR)
measurements.\cite{kasai} This value of $\it{g^{*}}$ is lower than
that measured in the bulk sample D ($\it{g^{*}}$ = 2.03), which
may be attributed to the in-plane compressive strain in the
epilayers due to the lattice mismatch with the substrate. Figure
1(c) shows ${\it T}_{2}^{*}$ decreasing by almost 60$\%$ when the
magnetic field is increased from $\it{B}$ = 0.l T to 1 T for
samples A and B, and by about 30$\%$ for sample C. This field
dependence is similar to that observed in GaN.\cite{beschoten}\par

\begin{figure}
\includegraphics[width=0.44\textwidth]{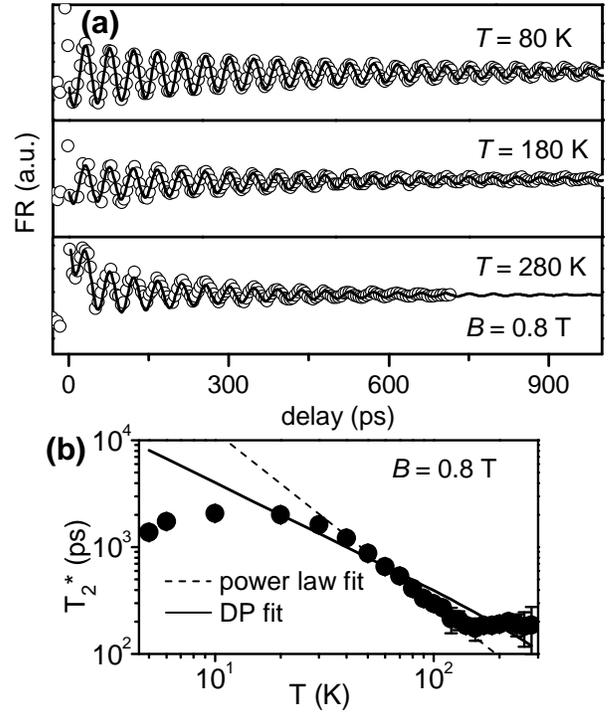}
\caption{\label{fig:epsart}Temperature dependence of TRFR on
sample A. (a) Spin precession at three temperatures for $\it{B}$ =
0.8 T. (b) $\it{T}_{2}^{*}$ follows a power law (dashed line) with
temperature in the intermediate region of 30 to 150 K, with an
exponent -1.54 $\pm$  0.03. Theoretical fit to the data for DP
mechanism is shown as a solid line.}
\end{figure}

 In Fig. 2, we perform a detailed study of the temperature
 dependence of ${\it T}_{2}^{*}$ in sample A, which has the highest carrier
 density, and the longest spin coherence time of all the epilayers.
  Figure 2(a) shows the TRFR response at $\it{B}$ = 0.8 T for temperatures
  $\it{T}$ = 80 K, 180 K and 280 K. The g-factor $\it{g^{*}}$ is temperature independent,
   while ${\it T}_{2}^{*}$ decreases with increasing $\it{T}$. Remarkably, spin coherence
   persists until room temperature, with a spin lifetime of 188 ps
   at 280 K, considerably longer than the spin coherence time
   observed in GaN ($\sim$35 ps).\cite{beschoten} Figure 2(b) is a logarithmic
   plot of $\it{T_{2}^{*}}$ obtained from TRFR data at $\it{B}$ = 0.8 T, as a function
   of $\it{T}$. For 5.5 K $\le$ $\it{T}$ $\le$ 30 K, $\it{T_{2}^{*}}$ is weakly temperature dependent,
   remaining almost constant at 2 ns; for 30 K $\le$ $\it{T}$ $\le$ 150 K, it
   follows a power law $\it{T^{-\alpha}}$, with $\alpha$ = 1.54 $\pm$ 0.03; for
   150 K $\le$ $\it{T}$ $\le$ 290 K, it is relatively flat, fluctuating around
   190 ps. \par

\begin{figure}
\includegraphics[width=0.44\textwidth]{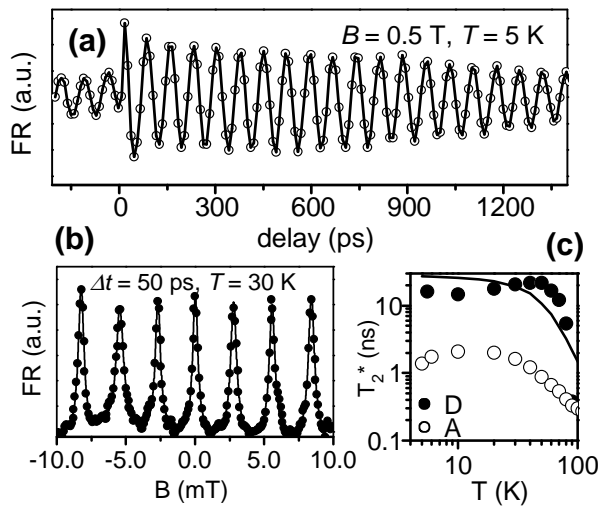}
\caption{\label{fig:epsart}(a) TRFR at $\it{T}$ = 5 K for the bulk
sample D. Line is a guide to the eye. (b) Faraday rotation versus
$\it{B}$ at $\it{T}$ = 30 K and $\Delta t$ = 50 ps on sample D
showing multiple RSA peaks. Lines are fits to data. (c)
$\it{T}_{2}^{*}$ obtained from RSA peak at $\it{B}$ = 0 T, as a
function of temperature for sample D (filled circles) with data
from sample A (open circles) shown for comparison. The line is a
theoretical fit of the DP mechanism for sample D.}
\end{figure}

TRFR data at $\it{T}$ = 5 K and $\it{B}$ = 0.5 T for the bulk
sample D is shown in Fig. 3(a). We observe that the Faraday signal
exhibits oscillations at negative delay, indicative of spin
magnetization persisting from the previous laser pulse, suggesting
$\it{T_{2}^{*}}$ $\geq$ $\it{t_{rep}}$. We employ the method of
Resonant Spin Amplification (RSA) which makes it possible to
measure spin coherence times well in excess of
$\it{t_{rep}}$.\cite{kikkawa} Here, the Faraday rotation is
measured at a fixed pump - probe delay. By varying the magnetic
field, and thus, the Larmor frequency, resonant enhancements in
Faraday rotation occur when the spin precession and pulse
repetition periods are commensurate, due to constructive
interference of successively excited spin packets.\par

Figure 3(b) shows multiple RSA peaks in a magnetic field scan from
+10 mT to -10 mT at  $\Delta t$ = 50 ps and $\it{T}$ = 30 K. From
a Lorentzian fit to the zero-field resonance peak, we obtain a
${\it T}_{2}^{*}$ of 20 ns. Figure 3(c) shows a log-log plot of
the temperature dependence of ${\it T}_{2}^{*}$ in the bulk sample
obtained from such fits for 5 K $\le$ $\it{T}$ $\le$ 100 K. For
comparison, we reproduce the temperature dependence of sample A
from Fig. 2(b), and notice that despite having carrier
concentration four orders in magnitude lower, ${\it T}_{2}^{*}$ is
larger by a factor of ten in the bulk. It also does not follow the
power law of ${\it T}^{-3/2}$ seen in the epilayer. We speculate
that these differences between the bulk and the epilayer may be
attributed to the presence of compressive strain and higher
density of defects in the latter.  \par

Of the three spin decoherence mechanisms relevant in
semiconductors -  Elliot-Yafet (EY),\cite{elliot} D'yakonov-Perel
(DP),\cite{DP} and Bir-Aronov-Pikus (BAP) \cite{BAP} mechanisms,
the EY process should not be very efficient in ZnO due to its
large band-gap and small spin-orbit splitting,\cite{madelung}
 while the BAP process should only make a significant
contribution to the spin relaxation rate when the concentration of
holes is high ($\sim$10$^{17}$ cm$^{-3}$). We show in Fig. 2(b)
and 3(c) (solid lines) the electron spin relaxation rate for the
DP process, calculated using

\begin{equation}
[{\it T}_{2}^{-1}]_{DP} =
\frac{24m_{e}^{*}\gamma_{1}^{2}k_{B}T\tau_{tr}}{{\hbar^{4}}} +
\frac{256(m_{e}^{*})^{3}\gamma_{3}^{2}E_{g}(k_{B}T)^{3}\tau_{tr}}{21\hbar^{6}}.
\end{equation}

where\cite{list} m$_{e}^{*}$ is the electron effective mass,
E$_{g}$ the band gap, $\gamma_{1}$ and $\gamma_{3}$ the
spin-splitting coefficients and $\tau_{tr}$ is the transport time,
related to the mobility $\mu$ via $\tau_{tr}= \mu m_{e}^{*}/{q}$
(q is the electron charge). While it predicts the correct
qualitative trends, it does not explain all the data, especially
the persisting spin coherence at high temperatures, suggesting
that there may be other spin scattering processes responsible for
our findings.\par

In conclusion, we have studied spin coherence in bulk and epilayer
samples of ZnO. In the bulk sample, $\it{T_{2}^{*}}$ extends to
very long times at low temperatures. In the epilayers, ${\it
T}_{2}^{*}$ increases with carrier concentration, but is much
smaller than in the bulk. However, spin precession in the
epilayers persist until room temperature, adding to the
attractiveness of ZnO as a material for spintronics.\par

We thank R. J. Epstein and Y. K. Kato for discussions and AFOSR,
ARO, ONR and DARPA/DMEA for financial support.

%\newpage

%\begin{center}{\bf Figure Captions}\end{center}


\begin{thebibliography}{16}


\bibitem{deitl}T. Deitl, H. Ohno, F. Matsukura, J. Cibert, and D. Ferrand, Science {\bf287}, 1019 (2000).
\bibitem{wolf} S. A. Wolf, D. D. Awschalom, R. A. Buhrman, J. M. Daughton, M. L. Von Molnar, S. Roukes, A. Y. Chtchelkanova and D. M. Treger, Science {\bf294}, 1488 (2001).
\bibitem{pearton} S. J. Pearton, D. P. Norton, K. Ip, Y. W. Heo, and T. Steiner, J. Vac. Sci. Technol. B {\bf22}(3), 932 (2004).
\bibitem{jung} S. W. Jung, S. J. An, G. C. Yi, C. U. Jung, S. I. Lee, and S. Cho, Appl. Phys. Lett. {\bf80}, 4561 (2002).
\bibitem{fukumura} T. Fukumura, Z. Jin, M. Kawasaki, T. Shono, T. Hasegawa, S. Koshihara, and H. Koinuma, Appl. Phys. Lett. {\bf78}, 958 (2001).

\bibitem{look} D. C. Look, J. W. Hemsky, and J. R. Sizelove, Phys. Rev. Lett. {\bf82}, 2552
(1999).
\bibitem{chris} C. G. Van de Walle, Phys. Rev. Lett. {\bf85}, 1012 (2000).
\bibitem{ueda} K. Ueda, H. Tabata, and T. Kawai, Appl. Phys. Lett. {\bf79}, 988
(2001).
\bibitem{minegishi} K. Minegishi, Y. Koiwai, Y. Kikuchi, K. Yano, M. Kasuga, and A. Shimizu, Japan. J. Appl. Phys. {\bf36}, L1453 (1997).
\bibitem{vaithianathan} V. Vaithianathan, B. T. Lee, and S. S. kim, Appl. Phys. Lett. {\bf86}, 062101 (2005).
\bibitem{beschoten} B. Beschoten, E. Johnston-Halperin, D. K. Young, M. Poggio, J. E. Grimaldi, S. Keller, S. P. DenBaars, U. K. Mishra, E. L. Hu, and D. D. Awschalom, Phys. Rev. B {\bf63}, 121202(R) (2001).
\bibitem{sakagami} N. Sakagami and K. Shibayama, Japan. J. of App. Phys. {\bf20}, 201 (1981).
\bibitem{crooker1}
S. A. Crooker, D. D. Awschalom, J. J. Baumberg, F. Flack and N.
Samarth, Phys. Rev. B {\bf56}, 7574 (1997).
\bibitem{crooker2} S. A. Crooker et al., Phys. Rev. Lett. {\bf77}, 2814
(1996).
\bibitem{kasai} P. Kasai, Physical Review {\bf130}, 989 (1963).
\bibitem{kikkawa} J. M. Kikkawa and D. D. Awschalom, Phys. Rev. Lett. {\bf80}, 4313 (1998).
\bibitem{elliot} R. J. Elliot, Phys. Rev. {\bf96}, 266 (1954).
\bibitem{DP} M. I. D'yakanov and V. I. Perel', Sov. Phys. JETP {\bf42}, 7905 (1976).
\bibitem{BAP} G. L. Bir, A. G. Aronov, and G. E. Pikus, Sov. Phys. JETP {\bf42}, 705
(1976).
\bibitem{madelung} O. Madelung,  "Semiconductors: Data Handbook", Springer
2003.
\bibitem{list} The parameters used in the calculations for [${\it T}_{2}$]$_{DP}$ in the epilayers
are $\gamma_{1}$ = 0.005 eV$\AA$, $\gamma_{3}$ = 5 eV$\AA^{3}$,
$\it{m^{*}_{e}}$=0.275 $\it{m_{e}}$, and $\it{E_{g}}$=3.445 eV. In
the epilayer, the measured mobility is temperature independent and
for the calculation we use
   $\mu$(20K-280K)=19 cm$^{2}$/Vs, for evaluating $\tau_{tr}$.
  In the bulk we use the measured values for the temperature-dependent
  mobility $\mu$, and tune $\gamma_{1}$ to 0.5 meV$\AA$, to account for the decreased strain compared
  to the epilayers.
\end{thebibliography}
\end{document}